\documentclass[%
 reprint,
 amsmath,amssymb,
 aps,
]{revtex4-2}

\usepackage{graphicx}
\usepackage{amsmath}
\usepackage{subfigure}
\usepackage{dcolumn}
\usepackage{bm}
\usepackage{appendix}

\begin{document}

\preprint{APS/123-QED}

\title{New $\gamma$-softness in  $^{196}$Pt from $\gamma$-rigid triaxiality }

\author{Chunxiao Zhou$^{1,2}$}
\email{zhouchunxiao567@163.com}
\author{Xue Shang$^{1,2}$}
\author{Tao Wang$^{3}$}%
 \email{suiyueqiaoqiao@163.com}
\affiliation{%
$^{1}$College of Mathematics and Physics Science, Hunan University of Arts and Science, Changde 415000, People's Republic of China\\
$^{2}$Hunan Province Key Laboratory of Optoelectronic Information Integration and Optical Manufacturing Technology, Changde, 415000, China\\
$^{3}$College of Physics and Engineering Technology, Tonghua Normal University, Tonghua 134000, People's Republic of China
}%

\date{\today}

\begin{abstract}
The interplay between the $\gamma$-softness and the $\gamma$-rigidity is a key aspect in understanding nuclear collectivity. In this paper, the $\gamma$-softness emerging from the $\gamma$-rigid triaxiality within the framework of the SU3-IBM is applied to investigate the $^{196}$Pt nucleus. The SU(3) fourth-order interaction, $\hat{C}_{2}^{2}[\text{SU(3)}]$, which introduces the rigid triaxial degree of freedom, is treated as an essential component of the Hamiltonian. Based on this, we systematically calculate the low-lying energy spectra, B(E2) transition rates, and quadrupole moments for $^{196}$Pt. The agreement between experimental data and theoretical calculations demonstrates that the present approach provides an effective explanation for the observed features.

\end{abstract}

\maketitle


\section{INTRODUCTION}

 Nuclear shape is one of the fundamental concepts in nuclear structure,  as it reflects the long-range correlations among the nucleons \cite{Bohr1975NuclearSV,RevModPhys.82.2155,heyde2016nuclear}. The interacting boson model (IBM), proposed by Arima and Iachello \cite{PhysRevLett.35.1069,ARIMA1976253,ARIMA1978201,SCHOLTEN1978325,ARIMA1979468}, provides an elegant algebraic framework for describing nuclear shapes. In its simplest version, IBM-1 \cite{Iachello_Arima_1987}, the nucleus is described in terms of $s$ ($L=0$) and $d$ ($L=2$) bosons representing nucleon pairs. The corresponding Hamiltonian contains only up to two-body interactions and classifies nuclear shapes into four symmetry limits: spherical (the U(5) symmetry limit), prolate (the SU(3) symmetry limit), oblate (the $\overline{\text{SU(3)}}$ symmetry limit) and $\gamma$-soft (the O(6) symmetry limit) shapes. Within this framework, various shape phase transitions have been investigated \cite{RevModPhys.82.2155,casten2006shape,Casten_2007,CASTEN2009183,CEJNAR2009210,FORTUNATO2021103891,jolos2021phase,PhysRevC.68.034326,Pan_2008,PhysRevC.80.014311,PhysRevC.85.054309}, including the prolate-oblate shape phase transition \cite{PhysRevLett.87.162501,PhysRevC.68.031301}. However, when describing the rigid triaxial shape, it is necessary to go beyond the two-body interactions and to incorporate the higher-order interactions (e.g., the $6-d$ interaction $[d^{\dag} d^{\dag} d^{\dag}]^{(L)}\cdot[\tilde{d}\tilde{d}\tilde{d}]^{(L)}$ \cite{PhysRevC.24.684, PhysRevC.29.1420} or the SU(3) higher-order interactions \cite{PhysRevC.61.041302}).

While the IBM-1 provides an algebraic language for describing quadrupole collectivity, its quantitative description is inadequate for certain experimental phenomena, such as the B(E2) anomaly \cite{PhysRevC.94.044327,PhysRevC.96.021301,PhysRevLett.121.022502,PhysRevC.100.034302} and the Cd puzzle \cite{PhysRevC.78.044307,PhysRevC.86.044304,PhysRevC.86.064311,PhysRevLett.123.142502,PhysRevC.101.044302}. These experimental observations challenge the conventional IBM paradigms and motivate the new development of the model. Recently, an extended version of the IBM has been proposed (SU3-IBM), which includes the U(5) symmetry limit and the SU(3) symmetry limit with higher-order interactions \cite{Wang_2020,Wang_2022,universe12040091}. This new model unifies the prolate, oblate, and triaxial quadrupole deformations within the SU(3) symmetry, offering an intuitive view of nuclear structure. The SU3-IBM has successfully explained the B(E2) anomaly \cite{Wang_2020,ZHANG2022137443,PhysRevC.107.064303,PhysRevC.110.054324, Cheng25,Zhang253,Jin25} and provided a compelling explanation for the $\gamma$-soft spherical-like spectra in the Cd isotopes \cite{Wang_2022,universe12040091,10.1088/1674-1137/ad7d76,WangPd2025,Zhao26}. Additionally, it offers insightful interpretations of several other notable phenomena, such as the prolate-oblate shape asymmetric transitions in the Hf-Hg region \cite{PhysRevC.84.014326,Zhang2012,PhysRevC.107.064322,Teng251}, and the E(5)-like spectra in $^{82}$Kr \cite{PhysRevC.108.024309}. Most recently, this model has also been extended to describe the phenomenon of small rigid triaxiality in well-deformed nuclei such as $^{154}$Sm \cite{ZHOU2026140286} and $^{166}$Er \cite{prcsubmit2026}.  

In the conventional IBM framework, the oblate shape is described by the $\overline{\text{SU(3)}}$ symmetry limit and the O(6) symmetry limit is regarded as the critical point of the phase transition from the prolate shape (the SU(3) symmetry limit) to the oblate shape (the $\overline{\text{SU(3)}}$ symmetry limit) \cite{PhysRevLett.87.162501,PhysRevLett.89.182502,nature2002,PhysRevC.68.031301,Bonatsos_2024}. In the SU3-IBM, by contrast, the oblate shape is described by the SU(3) third-order interaction \cite{PhysRevC.84.014326,Zhang2012}, which is distinct from the $\overline{\text{SU(3)}}$ picture, and a new type of $\gamma$-soft analogous to the O(6) symmetry limit was found to emerge at the middle location from the U(5) symmetry limit to the SU(3) degeneracy point appearing from the prolate shape to the oblate shape \cite{Wang_2022}. 

This description has been applied to study the properties of $^{196}$Pt \cite{Wang_2024}, but it overestimates the energy of the $0_{3}^{+}$ state. The origin of this discrepancy can be understood from the phase structure of the model. In Ref.~\cite{PhysRevC.84.014326}, a narrow shallow triaxial region was found between the prolate and oblate phases. The parameters adopted in Ref.~\cite{Wang_2022} lie close to the prolate-oblate degeneracy point, where a triaxial character is therefore expected to develop. However, the tiny triaxiality generated in this way cannot account for the observed properties of  $^{196}$Pt. We therefore incorporated a small SU(3) fourth-order interaction \cite{Wang_2024}, which introduces rigidity ro the triaxial region and brings the $0_{3}^{+}$ state energy into agreement with experiment. While this prescription resolves the discrepancy, the underlying physical mechanism remains insufficiently understood.

To clarify this issue, we revisit the $\gamma$-soft behavior of  $^{196}$Pt in this work. In contrast to the treatment of Ref.~\cite{Wang_2024}, where the fourth-order interaction was introduced as a minor term, we include it as an essential component of the Hamiltonian. This treatment is motivated by some new studies, in which traditionally $\gamma$-soft nuclei can be described by the rigid triaxial model. For example, in finite boson systems, the equivalence between a $\gamma$-unstable description and a rigid triaxial description (with $\gamma$=30$^\circ$) has been demonstrated using the O(6) symmetry limit for $^{196}$Pt \cite{PhysRevLett.59.1541,PhysRevC.38.1038}. Furthermore, the rigid triaxial rotor model has also been applied to describe the $\gamma$-soft nuclei, such as $^{186,188,190,192}$Os and $^{196}$Pt \cite{PhysRevC.70.024308, PhysRevC.78.014302, PhysRevC.81.051305}. 
Building on this observation, the $\gamma$-softness in this study is built upon a foundation of rigid triaxiality. By combining the U(5) symmetry limit with the triaxial rotation of the SU(3) symmetry limit with higher-order interactions, we successfully reproduce the O(6)-like $\gamma$-soft feature. Comparisons with different models show that our approach reproduces experimental observables of $^{196}$Pt very well, which provides new insights into its new $\gamma$-softness.


The paper is organized as follows. Section II presents the Hamiltonian used in this work. In Section III, we compare the traditional $\gamma$-soft behavior of the O(6) symmetry limit with that emerging from the SU3-IBM framework. Section IV displays the calculated excitation spectra, B(E2) transition rates, and quadrupole moments. A detailed discussion of the new $\gamma$-softness in the present context is provided in Section V. Finally, Section VI summarizes the main results and conclusions.

\section{HAMILTONIAN}

The Hamiltonian that describes the evolution from the U(5) symmetry limit to the O(6) symmetry limit can be expressed as 
\begin{equation}\label{1}
\hat{H}_{0}=c_{0}\left[(1-\eta_{0})\hat{n}_{d}-\eta_{0} \frac{\hat{Q}_{0}\cdot\hat{Q}_{0}}{N}\right],
\end{equation}
where $c_{0}$ is the total fitting parameter, $0\leq\eta_{0}\leq1$. $\hat{n}_{d}$ is the $d$-boson number operator of the U(5) symmetry limit. $\hat{Q}_{0}$ is the quadrupole  operator of the O(6) symmetry limit. $\eta_{0}=0$ describes the spherical shape of the U(5) symmetry limit, while $\eta_{0}=1$ corresponds to the $\gamma$-softness of the O(6) symmetry limit. The transitional behavior of the low-lying states from the U(5) symmetry limit to the O(6) symmetry limit will be presented using the Hamiltonian (\ref{1}),  and the results will be compared with those from the SU3-IBM.

The Hamiltonian utilized in this paper can be expressed as
\begin{equation}\label{2}
\hat{H}=c\left[\hat{n}_{d}+\eta \hat{H}_{SU(3)}\right],
\end{equation}
where $c$ is the total fitting parameter and $\eta$ is the fitting parameter $\hat{H}_{SU(3)}$. The second part comprises a range of symmetry-conserving interactions in the SU(3) symmetry limit, which can be expressed as
\begin{equation}\label{3}
  \hat{H}_{SU(3)}=-\frac{\hat{C}_{2}[SU(3)]}{2N}+\kappa\frac{\hat{C}_{3}[SU(3)]}{2N^{2}}+\xi\frac{\hat{C}_{2}^{2}[SU(3)]}{2N^{3}},
\end{equation}
where $\kappa$ and $\xi$ are the coefficients of the third-order and fourth-order interactions, respectively, and $N$ is the boson number.
The two SU(3) Casimir operators are defined as
\begin{equation}\label{4}
 \hat{C}_{2}[SU(3)]=2\hat{Q}\cdot\hat{Q}+\frac{3}{4}\hat{L}\cdot\hat{L},
\end{equation}
\begin{equation}\label{5}
 \hat{C}_{3}[SU(3)]=-\frac{4\sqrt{35}}{9}[\hat{Q}\times\hat{Q}\times\hat{Q}]^{0}-\frac{\sqrt{15}}{2}[\hat{L}\times\hat{Q}\times\hat{L}]^{0},
 \end{equation}
where $\hat{Q}=[d^{\dag}\times\tilde{s}+s^{\dag}\times \tilde{d}]^{(2)}-\sqrt{7}/2[d^{\dag}\times \tilde{d}]^{(2)} $ is the quadrupole momentum operator of the SU(3) symmetry limit and $\hat{L}=\sqrt{10}[d^{\dag}\times\tilde{d}]^{(1)}$ is the angular moment operator. Under the group chain $\text{U(6)}\supset \text{SU(3)}\supset \text{O(3)}$, the eigenvalues of the Casimir operators can be expressed in terms of the SU(3) irreps ($\lambda$, $\mu$) as
\begin{equation}\label{6}
 \langle\hat{C}_{2}[SU(3)]\rangle=\lambda^{2}+\mu^{2}+\lambda\mu+3\lambda+3\mu,
\end{equation}
\begin{equation}\label{7}
 \langle\hat{C}_{3}[SU(3)]\rangle=\frac{1}{9}(\lambda-\mu)(2\lambda+\mu+3)(\lambda+2\mu+3).
\end{equation}

\begin{figure}[ht]
\includegraphics[width=0.9\columnwidth]{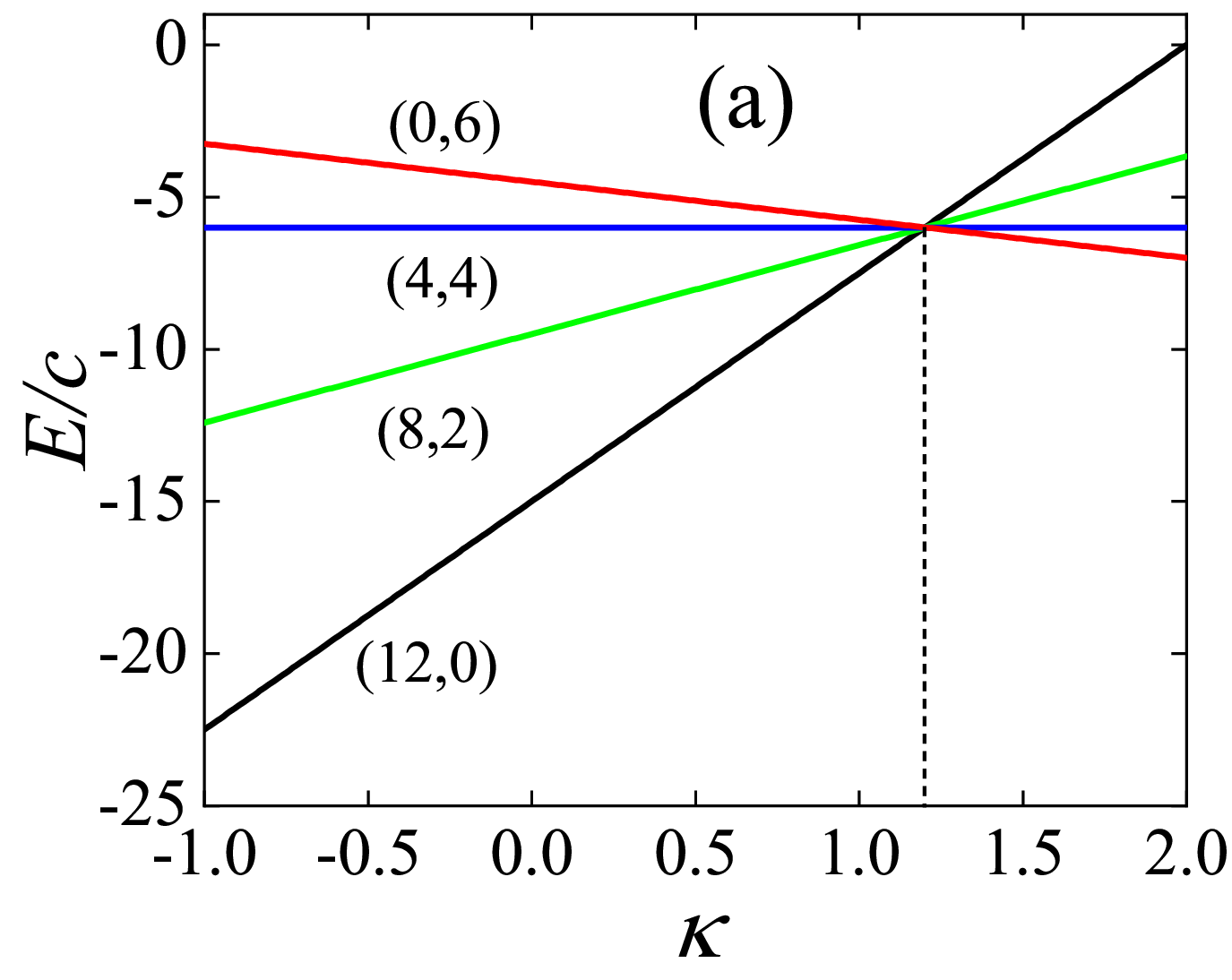}
\includegraphics[width=0.9\columnwidth]{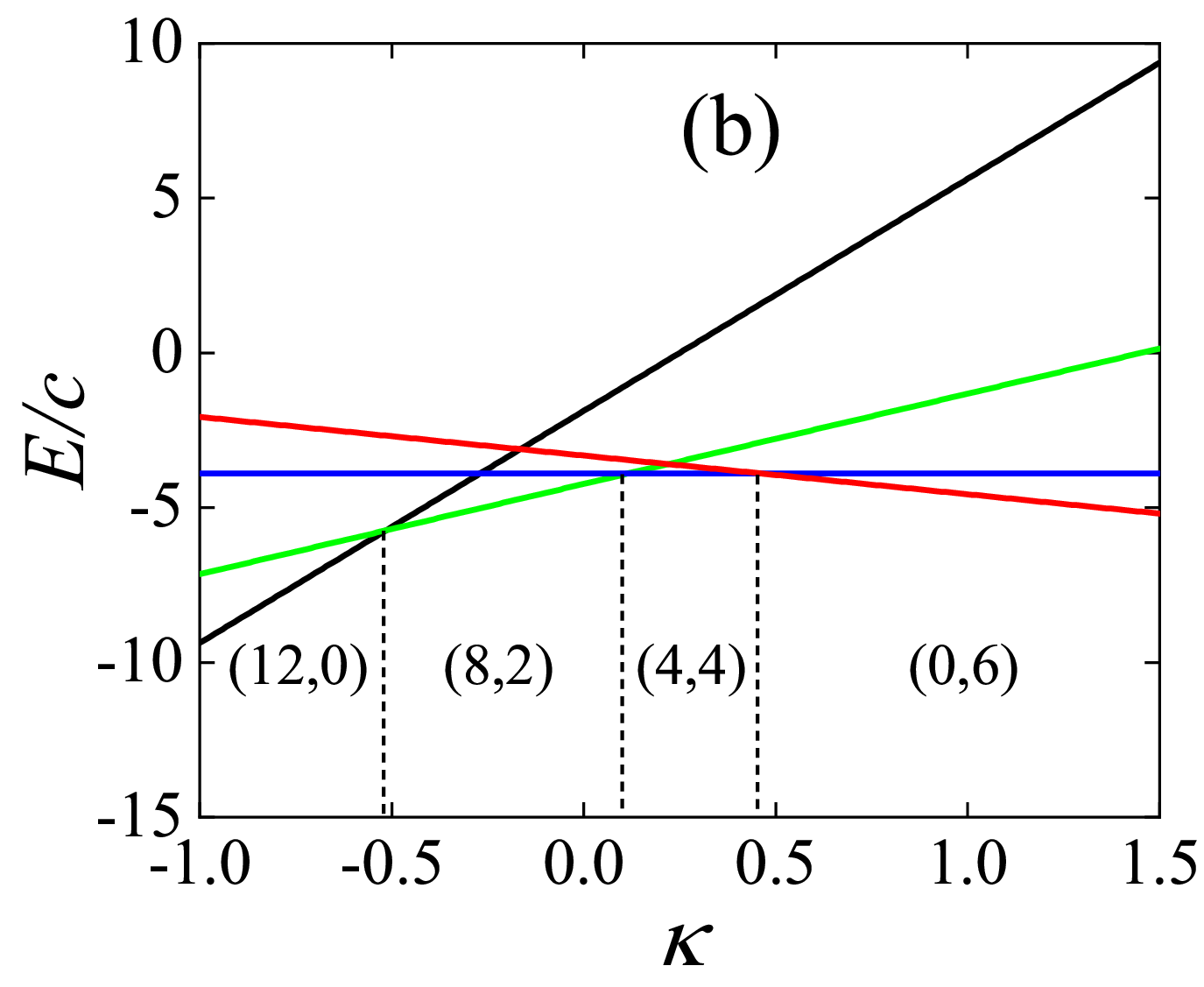}

\caption{(color online) Evolutional behaviors of the SU(3) irreps in the SU(3) symmetry limit for $N=6$. (a) Considering the second- and third-order interactions as a function of $\kappa$ with $\xi=0$. (b) Considering the second-, third- and fourth-order interactions as a function of $\kappa$ with $\xi=0.175$.
}\label{fig:3}
\end{figure}

In Fig. \ref{fig:3}, the evolution of SU(3) irreps in the SU(3) symmetry limit for $N=6$ is plotted according to Eq. (\ref{3}). When only second- and third-order interactions are considered, the $0^{+}$ states with $2\lambda+\mu=2N$ cross over at $\kappa=1.2$, as shown in Fig. \ref{fig:3}(a). For $\kappa<1.2$, the lowest irrep is (12,0), which corresponds to the prolate shape. However, for $\kappa>1.2$, the lowest irrep shifts to (0,6), corresponding to the oblate shape. 

The emerging $\gamma$-softness between the U(5) symmetry limit to the SU(3) degenerate point $\kappa=1.2$ has been demonstrated \cite{Wang_2022}, where the $\gamma$-softness was attributed to competition between the spherical, prolate and oblate shapes. When the four-body interaction is further taken into account, the degeneracy is lost, as depicted in Fig. \ref{fig:3}(b). It can be seen that not only the irreps (12,0) and (0,6) but also the irreps (8,2) and (4,4) can become the lowest. This incidats that the inclusion of the fourth-order interaction in the Hamiltonian enables the investigation of the rigid triaxial deformation \cite{PhysRevC.108.024309,ZHOU2026140286}, whereas it is not feasible when considering only up to third-order interaction. 

The nuclear shape can be characterized by the shape variables $(\beta,\gamma)$ of the collective model, which are related to the SU(3) irrep $(\lambda,\mu)$ \cite{Castanos1988,PhysRevC.90.044310}. The irrep $(\lambda, \mu)$ and the quadrupole shape variable $\gamma$ satisfy the following relation
\begin{equation}\label{8}
\gamma= \textrm{tan}^{-1} \left[\frac{\sqrt{3}(\mu+1)}{2\lambda+\mu+3}\right].
\end{equation} 

According to Eq. (\ref{8}), the irreps $(\lambda,\mu)$ corresponding to $\gamma=0$ and $\gamma=\pi/3$ 
describe to prolate and oblate shapes, respectively, while the intermediate value $0<\gamma<\pi/3$ describe various triaxial shapes. Any SU(3) irrep $(\lambda,\mu)$ can serve as the ground-state irrep of the system when the parameters $\kappa$ and $\xi$ in Eq. (\ref{3}) satisfy the following relationship \cite{PhysRevC.108.024309}

\begin{equation}\label{9}
\xi=\frac{N^{2}}{2g(\lambda,\mu)}-\frac{\kappa N}{6g(\lambda,\mu)}(3+\lambda+2\mu),
\end{equation}
where $g(\lambda,\mu)= \lambda^{2}+\mu^{2}+\lambda\mu+3\lambda+3\mu$.

It is important to note that, according to Eq. (\ref{9}), a one-to-one correspondence between $\kappa$ and $\xi$ for a given $(\lambda, \mu)$ holds strictly only in the large-$N$ limit. For finite $N$, this correspondence is relaxed. As shown in Fig. \ref{fig:3}(b) for $N=6$ and $\xi=0.175$, each irrep $(\lambda, \mu)$ corresponds to a range of $\kappa$. This broadening reflects the difference between the large-$N$ limit and the finite $N$ case, indicating that parameters derived from one regime cannot be directly transferred to another one.

\section{Analysis Of potential energy surface and transitional behaviors}

The main purpose of this paper is to demonstrate the emergence of the new O(6)-like $\gamma$-softness within the SU3-IBM framework. To elucidate this point, we compare the traditional $\gamma$-softness of the O(6) symmetry symmetry limit with that obtained from the SU3-IBM for both infinite and finite values of $N$. 
In the O(6) symmetry limit, the $\gamma$ degree of freedom is unrelated, with an average value of $\gamma_{eff}=30^{\circ}$. Our key finding is that an O(6)-like $\gamma$-soft regime can be generated by combining the $\hat{n}_{d}$ term with the rigid triaxial rotor. 

\subsection{Potential Energy Surface in the Large-$N$ imit}

\begin{figure}[ht]
 \includegraphics[width=0.8\columnwidth]{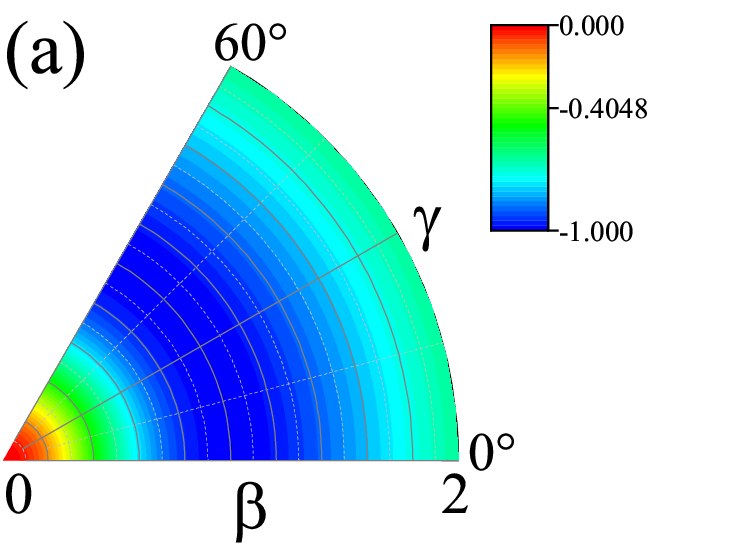}
\includegraphics[width=0.8\columnwidth]{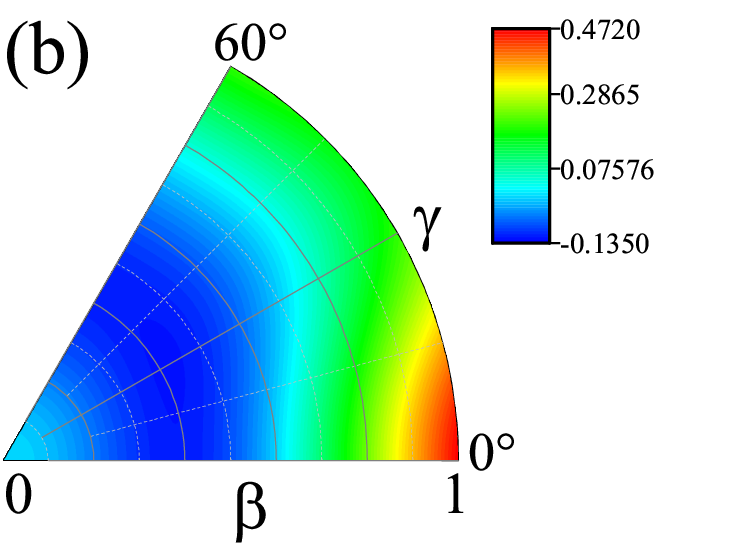}
\caption{(color online) Potential energy surfaces plotted in the large-$N$ limit for different models, (a) the O(6) symmetry limit, (b) the SU3-IBM with $\eta=0.701$, $\kappa=0.8$, $\xi = 0.175$.}\label{fig:4}
\end{figure}

The potential energy surface (PES) of the ground state in the large-$N$ limit provides a foundation for investigating nuclear shapes and testing structural models. As a reference for the $\gamma$-soft character, Fig. \ref{fig:4}(a) shows the PES of the O(6) symmetry limit, where a uniform minimum region in the $\gamma$ degree of freedom is centered around $\beta=1$. For the SU3-IBM considered here, the PES corresponding to the Hamiltonian (\ref{2}) can be expressed as follows\cite{PhysRevC.84.014326}

\begin{equation}
\begin{aligned}\label{10}
V(\beta,\gamma) = &\frac{\beta^{2}}{1+\beta^{2}} + \eta\bigg\{ 
    \frac{-\beta^{2}}{2(1+\beta^{2})^{2}} \big[8+\beta^{2}+4\sqrt{2}\beta\cos 3\gamma \big] \\
    &+ \frac{\xi \beta^{4}}{2(1+\beta^{2})^{4}}\big[64+32\beta^{2}+\beta^{4}+16\beta^{2} \cos 6\gamma \\
    &\quad + 8\sqrt{2}(8\beta+\beta^{3})\cos 3\gamma\big] \\
    &+ \frac{\kappa\beta^{3}}{9(1+\beta^{2})^{3}}\big[24\beta+16\sqrt{2}\cos 3\gamma \\
    &\quad + 6\sqrt{2}\beta^{2}\cos 3\gamma+\beta^{3}\cos 6\gamma\big]\bigg\}.
\end{aligned}
\end{equation}

In the large-$N$ limit, the irrep corresponding to $\gamma=30^{\circ}$ in the SU(3) limit is $(\frac{2}{3}N,\frac{2}{3}N)$.  Here, we set $\kappa = 0.8$, and according to Eq. (\ref{9}), $\xi = 0.175$. As shown in Fig. \ref{fig:4}(b), when $\eta=0.701$ the potential energy surfaces exhibit a broad minimum region around $\beta=0.4$ in the range $0^{\circ}<\gamma<60^{\circ}$. This characteristic is similar to that of the O(6) symmetry limit depicted in Fig. \ref{fig:4}(a), indicating that the SU3-IBM can effectively demonstrate the $\gamma$-softness in the large-$N$ limit.

\subsection{Transitional behavior of low-lying states for $N=6$}

\begin{figure}[ht]
\includegraphics[width=0.9\columnwidth]{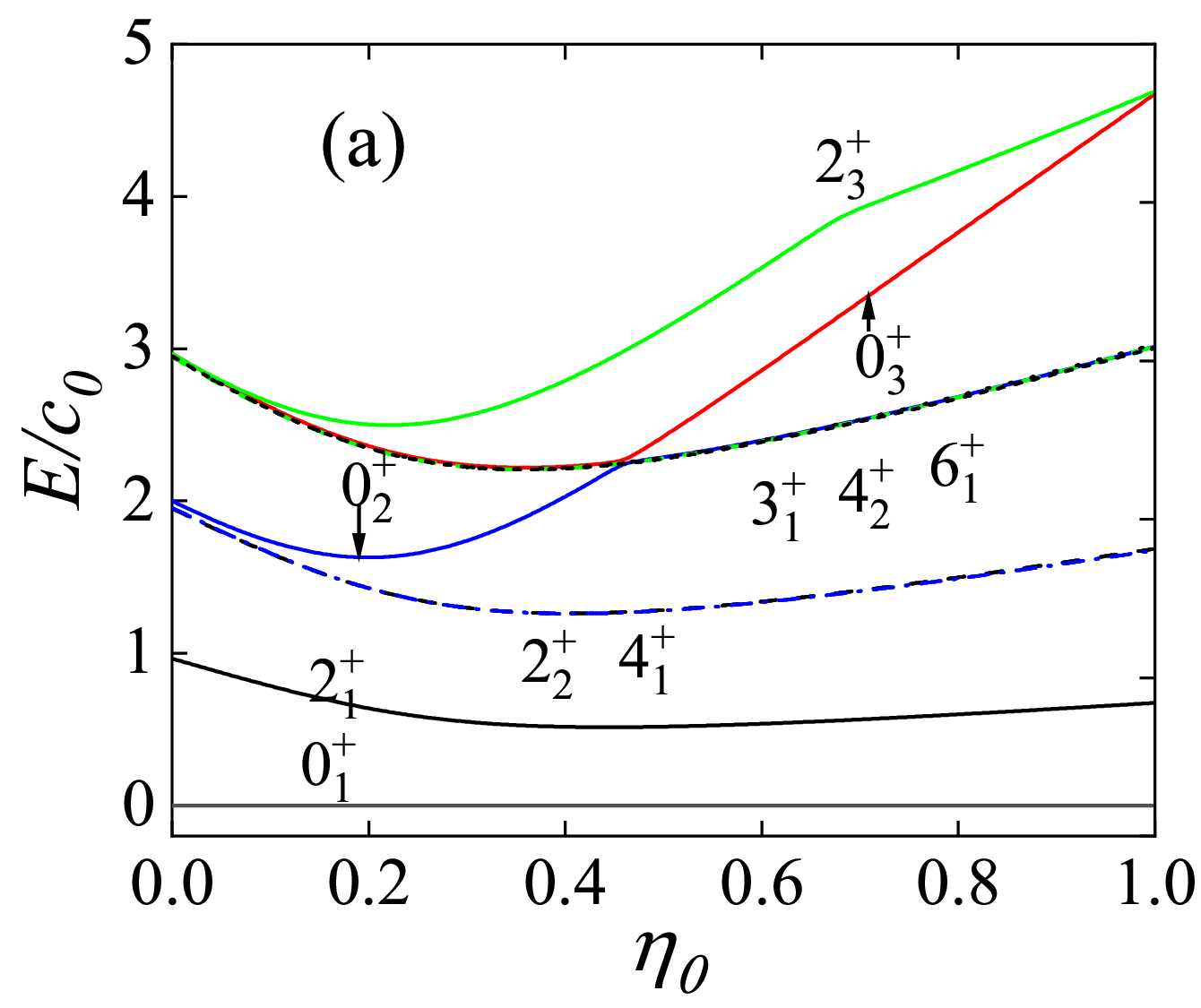}
\includegraphics[width=0.9\columnwidth]{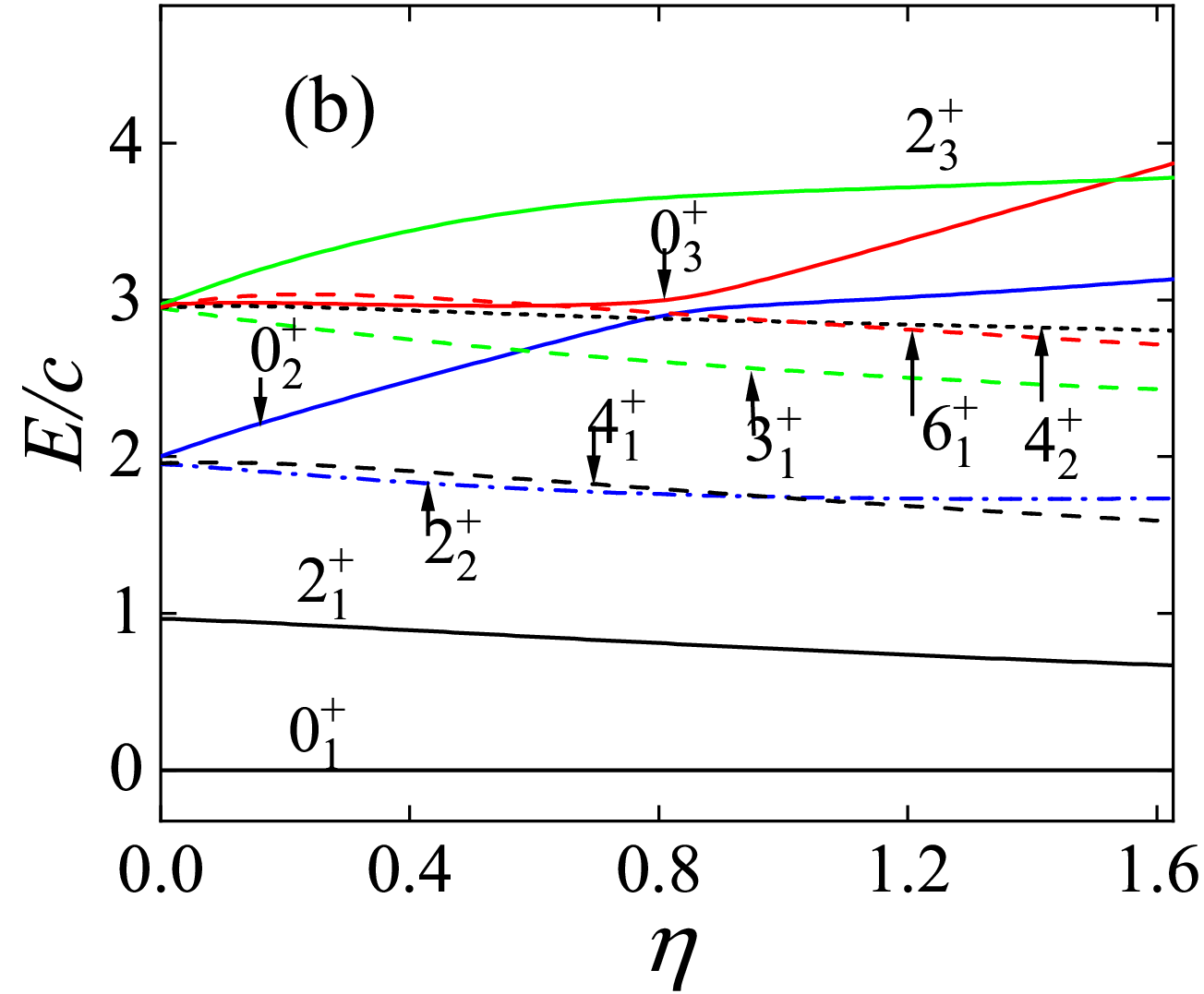}
\caption{(color online) Transitional behaviors of partial low-lying states as functions of the parameter $\eta$, (a) from the U(5) symmetry limit to the O(6) symmetry limit according to Eq. (\ref{1}), (b) the result obtained using Eq (\ref{2}) with $\kappa=0.47$ and $\xi=0.175$.
}\label{fig:5}
\end{figure}

To examine the new O(6)-like $\gamma$-softness of the proposed model for finite $N$, we analyze the transitional behaviors of low-lying states for the case of $N=6$. Fig. \ref{fig:5}(a) displays the transitions from the U(5) symmetry limit to the O(6) symmetry limit, where degeneracy is observed between the $2^{+}_{2}$ and $4^{+}_{1}$ states, as well as among the $3^{+}_{1}$, $4^{+}_{2}$ and $6^{+}_{1}$ states. A level crossing is also observed between the $0^{+}_{2}$ and $0^{+}_{3}$ states. In the SU3-IBM, we chose the SU(3) irrep (4,4) as the lowest $0^{+}$ state in the SU(3) symmetry limit (corresponding to $\gamma=30^{\circ}$), then set $\kappa=0.47$ and $\xi=0.175$. As shown in Fig. \ref{fig:5}(b), similar transitional behavior to that in Fig. \ref{fig:5}(a) can be observed when $\alpha=0.615$. Specifically, it captures the quasi-degenerate doublet $2^{+}_{2}$, $4^{+}_{1}$ and the near-degenerate triplet $3^{+}_{1}$, $4^{+}_{2}$, $6^{+}_{1}$, along with a strikingly similar evolution of the $0_{2}^{+}$ and $0_{3}^{+}$ states. These close agreements between Fig. \ref{fig:5}(a) and (b) demonstrate that the SU3-IBM can provides a capable framework for investigating the $\gamma$-soft feature of the realistic nulcei.

It is worth noting that the chosen parameters $\kappa=0.47$ and $\xi=0.175$ do not strictly satisfy the Eq. (\ref{9}), which can be attributed to two reasons. First, as shown in Fig. \ref{fig:3}(b), the lowest irrep (4,4)  is associated with a finite range of $\kappa$, rather than a single point. Second, Eq. (\ref{9}) is derived under the assumption of an exact SU(3) limit, whereas the $\hat{n}_{d}$ term in the Hamiltonian will breaks this symmetry and induces mixing among different SU(3) irreps.  As shown in Table I, the $0_{1}^{+}$ state does not correspond to a single irrep but to a combination of different irreps. Accordingly, the $\gamma$ value of ground state is not fixed to a specific value and should be interpreted as an effective mean, which can be calculated according to the relation \cite{PAN2003297}

\begin{equation}
\bar{\gamma}=\sum d_{(\lambda, \mu)}\gamma_{(\lambda, \mu)}, 
\label{eq11}
\end{equation}
where $d_{(\lambda,\mu)}$ denotes the decomposition probability of the SU(3) irrep $(\lambda,\mu)$ , and $\gamma_{(\lambda, \mu)}$ denotes the deformation parameter given in Eq. (\ref{8}). According to Eq. (\ref{eq11}) and Table. I, $\bar{\gamma}=32.5^{\circ}$ which is very close to the value $\gamma_{eff}=30^{\circ}$ of the O(6) symmetry limit. 

\section{THEORETICAL FITTING OF $^{196}$Pt}

\begin{table}
\renewcommand\arraystretch{1.3}
\caption{SU(3) decomposition of $0_{1}^{+}$ state. The parameters used are $\eta=1.63$, $\kappa=0.47$ and $\xi=0.175$.}
\begin{ruledtabular}
\begin{tabular}{l c c c c c c c c c}
\textrm{\quad}&
\textrm{(12,0)}&
\textrm{(8,2)}&
\textrm{(4,4)}&
\textrm{(0,6)}&
\textrm{(6,2)}&
\textrm{(2,4)}&
\textrm{(0,0)}\\
\colrule 
$d_{(\lambda,\mu)}$ & 0.001 & 0.116& 0.544& 0.200 & 0.067& 0.071 & 0.001 & 
\\ 
\end{tabular}
\end{ruledtabular}
\label{tab：Table1}
\end{table}


\begin{table}
\renewcommand\arraystretch{1.3}
\caption{\label{tab:table2}%
Absolute $B(E2)$ values in W.u. for $E2$ transitions between the low-lying normal states in $^{196}$Pt. The data of Experimental, PDS and MSR are taken from Ref. \cite{XIAOLONG20071093,PhysRevLett.102.112502,sym14122610}. Our results are calculated with $\eta=1.63$, $\kappa=0.47$, $\xi=0.175$ and $c=362.5$ keV.  The corresponding effective charge of
our calculation is $e$=2.551 (W.u)$^{1/2}$, respectively.
}

\begin{ruledtabular}
\begin{tabular}{cccccccc}
\textrm{$L_{i}$}&
\textrm{$L_{f}$}&
\textrm{Expt.}&
\textrm{PDS}&
\textrm{MSR}&
\textrm{Res.(I)}&
\textrm{Calc.}\\
\colrule
$2_{1}^{+}$ & $0_{1}^{+}$ & 40.60(20) & 40.6 & 40.6 & 40.6 & 40.6 \\

$2_{2}^{+}$ & $2_{1}^{+}$ & 54(+11 -12) & 53.0 & 45.3 & 43.2 &50.3\\
\quad & $0_{1}^{+}$ & $<7.8 \times 10^{-6}$ & 0.27 & 1.4 & 5.05 & 3.95\\

$4_{1}^{+}$ & $2_{1}^{+}$ & 60.0(9) & 53.0 & 53.6 & 51.4 & 51.9\\

$0_{2}^{+}$ & $2_{1}^{+}$ & 2.8(15) & 0.44 & 3.1 & 0.122  & 0.03\\
\quad & $2_{2}^{+}$ & 18(10) & 54.1 & 69.6 &79.5 & 58.3\\

$6_{1}^{+}$ & $4_{1}^{+}$ & 73(+4 -73) & 54.1 & 54.7 & 52.1  & 47.1\\

$4_{2}^{+}$ & $4_{1}^{+}$ & 17(6) & 25.8 & 18.1 & 27.4  & 27.8\\
\quad & $2_{1}^{+}$ & 0.56(+12 -17) & 0.14 & 0.69 & 1.34 & 1.06\\
\quad & $2_{2}^{+}$ & 29(+6 -29) & 28.3 & 28.8 & 31.2 &26.03\\

$2_{3}^{+}$ & $4_{1}^{+}$ & 0.13(12) & 0.059& 0.6 & 0.23 & 0.24\\
\quad & $2_{2}^{+}$ &0.26(23) & 0.19 & 0.27 & 3.47 & 2.71\\
\quad & $0_{1}^{+}$ &0.0025(24) & 0 & 0.064 & 0.02 & 0.03\\
\quad & $0_{2}^{+}$ &5(5) & 17.6 & 0.003 & 36.5 & 34.3\\

$0_{3}^{+}$ & $2_{1}^{+}$ & $<5.0$ & 0 & 1.05 & 6.82 & 8.26\\
\quad & $2_{2}^{+}$ & $<0.41$ & 0.41 & 0.009 & 0.114 & 2.07\\

$6_{2}^{+}$ & $6_{1}^{+}$ & 16(5) & 15.3 & 9.5 & 17.7 & 14.5\\
\quad & $4_{1}^{+}$ & 0.48(14) & 0.12 & 0.50 & 0.77 & 0.49\\
\quad & $4_{2}^{+}$ & 49(13) & 32.7 & 32.9 & 31.9 & 24.7\\

$8_{1}^{+}$ & $6_{1}^{+}$ & 78(+10 -78) & \quad & 47.6 & 29.3 & 31.6\\

\end{tabular}
\end{ruledtabular}
\end{table}

The reduced transitional rate B(E2) value provides one of the most insightful understandings of the collective modes in nuclei. A comparison between experimental data and theoretical calculations for $^{196}$Pt is presented in Table \ref{tab:table2}. The calculated results (in the last column), obtained with parameters $\eta=1.63$, $\kappa=0.47$, $\xi=0.175$ and $c=362.5$ keV, show generally good agreement with experimental data . However, noticeable deviations appear for several transitions among the low-lying states: the calculated $\text{B(E2};0_{2}^{+}\rightarrow 2_{1}^{+})$ is smaller than the experimental result, while $\text{B(E2};2_{3}^{+}\rightarrow 0_{2}^{+})$, $\text{B(E2};0_{3}^{+}\rightarrow 2_{2}^{+})$ and $\text{B(E2};0_{2}^{+}\rightarrow 2_{2}^{+})$ are overestimated. Here we would like to note that the near-zero value of $\text{B(E2};2_{2}^{+}\rightarrow0_{1}^{+})$ has been successfully explained by considering the interference effect between the mixing angle and the trixial shape angle in the triaxial rotor model \cite{PhysRevC.70.024308}. A bridge between the two descriptions is provided by a mapping of the triaxial rotor Hamiltonian onto the SU(3) symmetry limit of the IBM, and the resulting mapping terms have recently been shown to account for the B(E2) anomalies \cite{ZHANG2022137443}. Therefore, incorporating dynamical terms from the rotor Hamiltonian mapping (e.g., $[\hat{L}\times\hat{Q}\times\hat{L}]^{(0)}$,$[(\hat{L}\times\hat{Q})^{(1)}\times(\hat{L}\times\hat{Q})^{(1)}]^{(0)}$) is a promising method for improving the description of the deviated B(E2) values, which will be the focus of our subsequent works.

\begin{figure*}
\includegraphics[width=2.0\columnwidth]{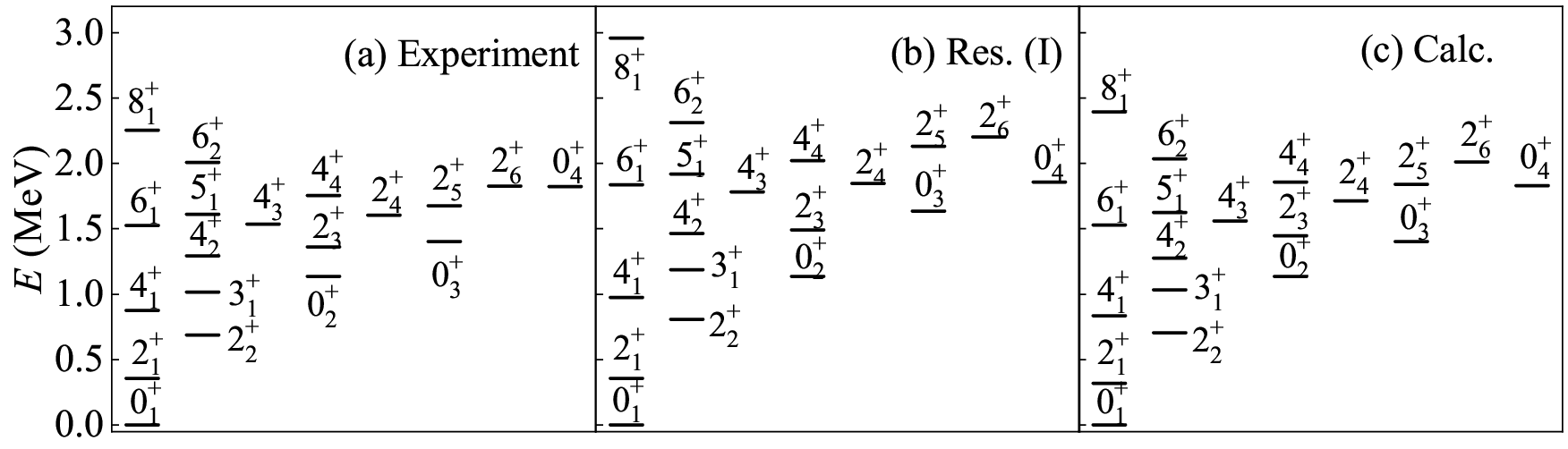}
\caption{Comparison of experimental and calculated partial low-lying level spectra for the $^{196}$Pt nucleus. (a) the experimental results, (b) the results from our previous work \cite{Wang_2024}, (c) the calculated results with $\eta=1.63$, $\kappa=0.47$, $\xi=0.175$ and $c=362.5$ keV (for better effect, the $L^{2}$ term is needed, whose parameter is 13.05 keV). The experimental data are taken from Ref. \cite{XIAOLONG20071093}.}\label{fig:7}
\end{figure*}

Despite some discrepancies in the B(E2) values, our model can excellently reproduce the experimental energy levels of $^{196}$Pt. A comparison is presented in Fig. \ref{fig:7}, where Fig. \ref{fig:7}(a) shows the experimental spectra. Fig. \ref{fig:7} (b) is the result from our previous work, in which the four-order interaction term serves as a supplementary term with a very small value. Fig. \ref{fig:7} (c) depict the present calculations with an enhanced treatment of the fourth-order interaction term, with $\xi=0.175$ (for better effect, the $L^{2}$ term is needed, whose parameter is 13.05 keV). As shown in Fig. \ref{fig:7}, our calculations agree well with the experimental results.  In particular, the states $8_{1}^{+}$, $6_{2}^{+}$ and $2_{6}^{+}$ in our model are lower than these obtained in Res. (I),  and are thus in closer agreement with the experimental energy spectra.

\begin{table}
 \renewcommand\arraystretch{1.3}
\caption{\label{tab:table3}%
Quadrupole moments in eb and average variance $\Delta Q$ of some low-lying states in $^{196}$Pt for various models. The data of Experimental, MSR and CQ are taken from Ref. \cite{XIAOLONG20071093,sym14122610}. The data of Res.(I) are taken from our previous work \cite{Wang_2024}. Our results are calculated with $\eta=1.63$, $\kappa=0.47$, $\xi=0.175$ and $c=362.5$ keV.
}

\begin{ruledtabular}
\begin{tabular}{cccccc}
\textrm{\quad}&
\textrm{Expt.}&
\textrm{MSR}&
\textrm{CQ}&
\textrm{Res.(I)}&
\textrm{Calc.}\\
\colrule
$Q(2_{1}^{+})$ & +0.62 & +0.50 & +0.40 &  +0.70 & +0.57  \\

$Q(2_{2}^{+})$ & -0.39 & -0.47 & -0.37 & -0.81 & -0.64\\

$Q(4_{1}^{+})$ & +1.03 & +0.66 & +0.40 & +0.63 & +0.48\\

$Q(6_{1}^{+})$ & -0.18 & +0.76 & +0.38 &  +0.35  & +0.18 \\

$\Delta Q$  & \quad & 0.51 & 0.44 &  0.39 & 0.35 \\

\end{tabular}
\end{ruledtabular}
\end{table}

In Table \ref{tab:table3}, the experimental and theoretical quadrupole moments of $^{196}$Pt are presented. It can be seen that the results given by different models all exhibit disparities with the experimental results (In the O(6) symmetry limit all the quadrupole moments are 0, which are not listed in Table \ref{tab:table3}). In particular, the experimental value of $Q(6_{1}^{+})$ is -0.18 eb, whereas none of the models presented yield a negative value. To quantify the overall agreement between the experimental and theoretical results, the root-mean-square deviation $\Delta Q$ is introduced. Our model gives the minimum value with 0.35 eb. These results indicate that our model provides the most accurate description of both the energy levels and quadrupole moments among the approaches compared.  In our subsequent works, we plan to incorporate the rigid-triaxial mapping terms as a mean to examine whether the discrepancy in $Q(6_{1}^{+})$ can be reduced.

\section{Some discussions on the new $\gamma$-softness in the SU3-IBM}


Some nuclei, such as Pt and Os isotopes, have level features that are different from the ones of the spherical shape and the prolate shape. This feature was first described by the $\gamma$-unstable model proposed by Wilets and Jeans \cite{PhysRev.102.788}. It exhibits a flat potential-energy surface in the $\gamma$ direction, and its collective wave functions of low-lying states are spread over a wide range of $\gamma$. The distribution of the ground state average over $\gamma$ can serve as a pivotal criterion for assessing the $\gamma$-softness. The O(6) symmetry limit of the IBM is known as a typical example of the $\gamma$-unstable system, where the ground state is based on an intrinsic state that is equally averaged over $0^{\circ} \leq \gamma \leq 60^{\circ}$. In traditional IBM, the O(6) symmetry limit is exactly solvable and the $\gamma$-unstable spectra can be predicted. The O(6) point is considered as the critical point of the phase transition from the SU(3) symmetry limit (prolate shape) to the $\overline{\textrm{SU(3)}}$ symmetry limit (oblate shape). The spectra of the $\overline{\textrm{SU(3)}}$ symmetry limit are the mirror image of the ones of the SU(3) symmetry limit, which implies the symmetric evolution from the prolate shape to the oblate shape. However, the symmetric shape evolution cannot be found in realistic nuclei, instead, it is found that the shape evolution is asymmetric. This reminded us that if the $\overline{\textrm{SU(3)}}$ cannot be achieved, the O(6) symmetry, despite its attractive mathematical form, cannot be easily achieved in realistic nuclei and was questioned in Ref. \cite{PhysRevC.107.064322,PhysRevC.107.064303}.

In the SU3-IBM, various quadrupole deformations are governed by the SU(3) symmetry. The oblate shape is  described by the third-order Casimir operator $\hat{C}_{3}$[SU(3)] interaction, which leads to asymmetric shape transition evolution between the prolate and oblate shapes. As shown in Fig. \ref{fig:3}(a), when considering the two- and three-body interactions, a degenerate point of the SU(3) irreps appears. With the inclusion of the $\hat{n}_{d}$ term, the new $\gamma$-softness emerges. However, our earlier calculations show that the energy spectra deviate significantly from the experimental data of $^{196}$Pt. This discrepancy was greatly removed by introducing a very small $\hat{C}_{2}^{2}$[SU(3)] term. As pointed out by Fortunato \textit{et al.} \cite{PhysRevC.84.014326}, a very tiny triaxial region exists during the transition from the prolate shape to the oblate shape, and the degeneracy coincides with this region. This suggests that the system at the degeneracy point  may not provide sufficient triaxial rigidity. The introduction of the $\hat{C}_{2}^{2}$[SU(3)] term enhances the triaxial stability of the system, yielding a more realistic energy spectra.

The distinction between the $\gamma$-rigid and $\gamma$-soft pictures is defined by fluctuations along the $\gamma$ degree of freedom. The rigid-triaxial rotor model corresponds to a deep potential minimum at a specific $\gamma$ value, indicating stable static triaxiality. In contrast, shallower minima signify the $\gamma$-softness. As noted in Ref. \cite{PhysRevLett.59.1541}, the soft and rigid triaxial description can be identical for the shallow minimum of the deformed potential. In this paper, we show that the $\gamma$-soft spectra can emerge from the $\gamma$-rigidity within the SU3-IBM. Using the parameters corresponding to the SU(3) irrep (4,4), our simulation for $^{196}$Pt reveals that the $\gamma$-softness arising from the rigid triaxiality exhibits better consistency with experimental data.

\section{CONCLUSIONS}

In this paper, we revisit the $\gamma$-soft behavior in $^{196}$Pt within the SU3-IBM. We show that the new O(6)-like $\gamma$-softness can emerge on the foundation of the $\gamma$-rigid triaxiality, which can be realized by combining the U(5) symmetry limit and the SU(3) symmetry limit. Unlike our previous paper, where the fourth-order interaction term serves as a supplementary term with very small value, this term plays a crucial role in the present model. Using this model, $B(E2)$ values, low-lying spectra and quadrupole moments are calculated. The calculated spectra and quadrupole moments agree well with the experimental data, whereas the remaining discrepancies in the B(E2) values indicate that further improvements are needed. The SU(3) mapping of the rigid triaxial rotor model will be incorporated to further elucidate the $\gamma$-softness in the following investigations, which is expected to improve the agreement between the calculated and the experimental results.

\begin{acknowledgments}

This work is supported by the National Natural Science Foundation of China (Grant No. 12104150) and Scientific Research Fund of Hunan Provincial Education Department (Grant No. 24B0639).
\end{acknowledgments}

\end{document}